# HARDWARE ARCHITECTURES FOR SUCCESSIVE CANCELLATION DECODING OF POLAR CODES

*Camille Leroux*[1], *Ido Tal*[2], *Alexander Vardy*[2], *Warren J. Gross*[1]

[1]McGill University
Montréal, Québec, Canada

[2]University of California San Diego
La Jolla, California, USA

**ABSTRACT**

The recently-discovered polar codes are widely seen as a major breakthrough in coding theory. These codes achieve the capacity of many important channels under successive cancellation decoding. Motivated by the rapid progress in the theory of polar codes, we propose a family of architectures for efficient hardware implementation of successive cancellation decoders. We show that such decoders can be implemented with $O(n)$ processing elements and $O(n)$ memory elements, while providing constant throughput. We also propose a technique for overlapping the decoding of several consecutive codewords, thereby achieving a significant speed-up factor. We furthermore show that successive cancellation decoding can be implemented in the logarithmic domain, thereby eliminating the multiplication and division operations and greatly reducing the complexity of each processing element.

## 1. INTRODUCTION

Polar codes [1] are a family of error correcting codes with an explicit construction and efficient encoding and decoding algorithms. Moreover, they achieve capacity (asymptotically in the code length $n$) if the underlying channel is symmetric, memoryless, and has binary-input. To date, no other family of codes possesses these attributes, hence polar codes are seen as a major breakthrough in coding theory. Not surprisingly, polar codes have garnered much interest recently in the coding theory community. From a practical point of view, the capacity of the channel can be approached at the expense of a large code length ($n = 2^{20}$ bits). In some information theoretic applications, polar codes are the only known solution which is both explicit and efficient: for example achieving the secrecy capacity of the wiretap channel in the general case [2]. Polar codes have recently been shown to have an efficient construction [3]. Recent results have started to address the issue of long code length. It was shown in [4] that applying belief propagation decoding on polar codes helps in reducing the required code length at the expense of extra complexity due to the iterative nature of belief propagation. Driven by the recent rapid progress in the theory of polar codes our motivation is to find efficient hardware architectures for SC decoding that will allow high-throughput and low-area implementations. Despite the numerous studies on polar codes construction and performance, the issue of hardware implementation of SC decoders remains an open problem. Initial results and a general framework for the implementation of belief propagation decoders for polar codes are given in [5]. However, due to its lower complexity compared to belief propagation, we are motivated to study the hardware implementation of successive cancellation decoding. Arıkan [1] showed that the SC decoding algorithm can be implemented in complexity $O(n \log_2 n)$, where $n$ is the code length.

In this paper, starting from the general framework proposed by Arıkan [1], we show that SC decoding can actually be implemented with hardware complexity $O(n)$. We also propose to increase the throughput by decoding several consecutive vectors at the same time. Finally, in order to reduce the complexity further, we address the implementation of the computational nodes by working in the logarithmic domain, thereby eliminating the multiplication and division operations. We show that the resulting transcendental functions can be approximated by the minimum function with negligible performance degradation.

## 2. POLAR CODES

Polar codes are linear block error-correcting codes. Assume from here onward that the underlying channel has binary input, is symmetric, and is memoryless. Fix $n = 2^m$ as the code length. Denote by $\mathbf{u} = (u_0, u_1, \ldots, u_{n-1})$ the input bits, and let $\mathbf{c} = (c_0, c_1, \ldots, c_{n-1})$ be the corresponding codeword[1]. The encoding operation has an FFT structure, depicted in Figure 1, for $m = 3$. Note that the ordering of the $u_i$ in Figure 1 is according to the bit-reversal order: if we reverse the order of the bits in the binary representation of $i$, then we get the standard lexicographic ordering.

Recall that $\mathbf{u}$ is encoded to $\mathbf{c}$. Next, $\mathbf{c}$ is sent over the underlying channel (the channel is used $n$ times). Denote by $\mathbf{y} = (y_0, y_1, \ldots, y_{n-1})$ the corresponding channel output. We now wish to decode $\mathbf{y}$. This is done in terms of a *successive cancellation* decoder. That is, given $\mathbf{y}$, we first try to deduce the value of $u_0$, then that of $u_1$, and so forth up until $u_{n-1}$. We do this as follows. Assume that we are currently at stage $i$, and so, we have already guessed the values of $u_0, u_1, \ldots, u_{i-1}$; denote these guesses as $\hat{u}_0, \hat{u}_1, \ldots, \hat{u}_{i-1}$. Next, for $b \in \{0, 1\}$, denote by $\Pr(\mathbf{y}|\hat{u}_0^{i-1}, u_i = b)$ the probability that $\mathbf{y}$ was transmitted, given that $u_0^{i-1} = \hat{u}_0^{i-1}$, that $u_i = b$, and that $u_{i+1}, u_{i+2}, \ldots, u_{n-1}$ are independent random variables with Bernoulli distribution of parameter $1/2$. If $i$ is not in the frozen set (explained later), then we take the guessed value $\hat{u}_i$ to be

$$\hat{u}_i = \begin{cases} 0, & \text{if } \frac{\Pr(\mathbf{y}|\hat{u}_0^{i-1}, u_i=0)}{\Pr(\mathbf{y}|\hat{u}_0^{i-1}, u_i=1)} > 1, \\ 1, & \text{otherwise,} \end{cases} \quad (1)$$

Consider the case in which we are at stage $i$, and $u_0^{i-1} = \hat{u}_0^{i-1}$ — that is, we have guessed correctly up until now. Then, as shown in [1], for almost all $0 \leq i < n$ we have that the probability of guessing $u_i$ correctly is either extremely close to 1 (very good), or extremely close to $1/2$ (very bad). That is, there is a *polarization* effect, as $n$ tends to infinity. In order to keep the assumption $u_0^{i-1} = \hat{u}_0^{i-1}$ valid for all $i$ (with very high probability), we freeze some $u_i$. That is, if the probability of guessing $u_i$ is not very good, then we set its value to 0 in both the encoder and the decoder, and thus no information is transmitted via $u_i$. As shown in [1], the fraction of indices $i$ which are not frozen (the effective code rate) tends to the capacity of the underlying channel.

---
[1]Note that $n$ input bits are encoded to a length $n$ codeword. However, as we will see later on, not all of the $n$ input bits carry information.

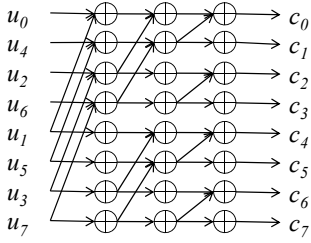

**Fig. 1**. Encoder architecture for $n = 8$

## 3. SUCCESSIVE CANCELLATION DECODER IMPLEMENTATION

### 3.1. FFT structure

Arıkan showed that SC decoding can be efficiently implemented by the factor graph of the code which has a structure resembling the Fast Fourier Transform (FFT). In the remainder of the paper, we will designate this decoder as the "FFT-like SC decoder". Figure 2 shows the graph of the SC decoder for $n = 8$. Channel likelihood ratios (LRs) $\lambda_i$ are assumed to be available on the right hand side of the graph while the estimated bits $\hat{u}_i$ are on the left hand side. The SC decoder is composed of $m = \log_2 n$ stages each containing $n$ nodes. We refer to a specific node as $\mathcal{N}_{l,j}$ where $l$ designates the stage index ($0 \leq l < m$) and $j$ designates the node index within stage $l$ ($0 \leq j < n$). Each node updates its output according to one of the two following update rules:

$$\begin{cases} f(a,b) = \frac{1+ab}{a+b} \text{ or,} \\ g_{\hat{u}_s}(a,b) = a^{1-2\hat{u}_s} b. \end{cases} \quad (2)$$

The values $a$ and $b$ are likelihood ratios while $\hat{u}_s$ is a bit that represents the partial modulo-2 sum of previously estimated bits. For example, in node $\mathcal{N}_{1,3}$, the partial sum is $\hat{u}_s = \hat{u}_4 \oplus \hat{u}_5$. The value of $\hat{u}_s$ determines if function $g$ should be a multiplication or a division. These update rules are complex to implement in hardware since they involve multiplications and divisions. In Section 4, we propose to perform these operations in the logarithmic domain and to apply an approximation to function $f$. For now we will consider $f$ and $g$ to be black boxes until we return to them in Section 4.

The sequential nature of the algorithm induces some data dependence within the processing. We notice that $\mathcal{N}_{1,2}$ can not be updated before the bit $\hat{u}_1$ is computed and a fortiori neither before $\hat{u}_0$ is known. In order to respect the data dependence, a scheduling has to be defined. Arıkan proposed two schedulings for this decoding framework [1]. In the left-to-right scheduling, nodes recursively call their predecessors until an updated node is reached. The recursive nature of this scheduling is especially suitable for software implementation. In the alternative right-to-left scheduling, any node updates its value whenever its inputs are available. Each bit $\hat{u}_i$ is successively estimated by activating the spanning tree rooted at $\mathcal{N}_{0,\pi(i)}$.

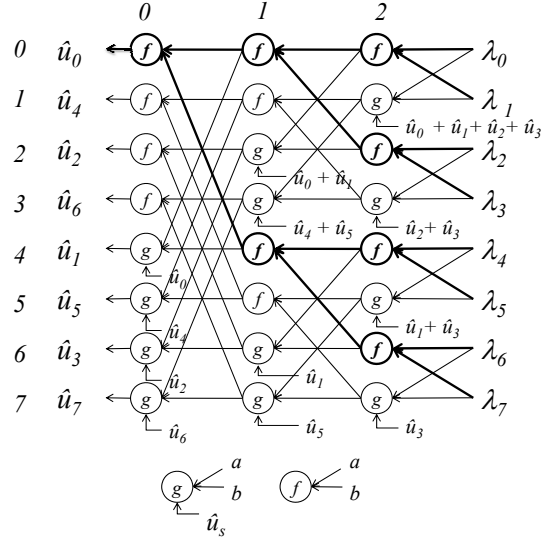

**Fig. 2**. FFT-like SC decoder architecture for $n = 8$

In Figure 2 the tree associated with $\hat{u}_0$ is highlighted. If we assume that a pipeline register is inserted between each stage or equivalently that each node processor can memorize its updated value, then some results can be reused. For example, in Figure 2, bit $\hat{u}_1$ can be decoded by only activating $\mathcal{N}_{0,4}$ since $\mathcal{N}_{1,0}$ and $\mathcal{N}_{1,4}$ have already been updated during the decoding of $\hat{u}_0$. Despite this well-defined structure and scheduling of the FFT-like decoder, in [1], Arıkan does not assess the problem of resource sharing, memory management or control generation that would be required for hardware implementation. This framework however suggests that it could be implemented with $n \log_2 n$ combinatorial node processors together with $n$ registers between each stage to memorize intermediate results. In order to store the channel information, $n$ extra registers are included as well. The total complexity of such a decoder is

$$C_T = (C_{\text{np}} + C_{\text{r}})n\log_2 n + nC_{\text{r}}, \quad (3)$$

where $C_{\text{np}}$ and $C_{\text{r}}$ are the hardware complexity of a node processor and a register respectively. It can be shown that such a decoder with the right-to-left scheduling would take $2n - 2$ clock cycles to decode $n$ bits. The throughput in bits per second would then be

$$T = \frac{n}{(2n-2)t_{\text{np}}} \approx \frac{1}{2t_{\text{np}}} \quad (4)$$

where $t_{\text{np}}$ is the propagation time in seconds through a node processor. It follows that every node processor is actually used once every $2n - 2$ clock cycles. This motivates us to find a schedule to merge some of the nodes into a single processing element.

### 3.2. Pipelined tree architecture

Looking further into the scheduling, we notice that whenever stage $l$ is activated, only $2^l$ nodes are actually updated. For example in Figure 2, when stage 0 is enabled, only one node is updated. Then the $n$ nodes of stage 0 can be implemented using a single processing element (PE). We note that in general, for stage $l$, $2^l$ processing elements (PEs) are sufficient to update the nodes. However, this resource sharing does not necessarily guarantee that the memories assigned to the merged nodes can also be merged. The memory sharing depends on the liveness of generated variables. Table 1 shows the stage activation during the decoding of one vector **y**. When stage $l$ is enabled, we indicate which function ($f$ or $g$) is applied to the $2^l$ activated nodes at stage $S_l$ during each clock cycle (CC). Every generated variable is used twice during the decoding. For example, the four variables generated in stage 2 at CC #1 are used on CC #2 and CC #5 in stage 1. This means that in stage 2, the four registers associated with the $f$ function can be reused at CC #8 to memorize the four data values generated by the $g$ function. This observation is applicable to any stage in the decoder. The resulting proposed architecture is shown in Figure 3 for $n = 8$. $n$ registers are used to store the LRs $\lambda_i$. The decoder is composed of a pipelined tree structure that includes $n - 1$ PEs, $P_{l,j}$, and $n - 1$ registers, $R_{l,j}$ with $0 \leq l \leq m - 1$ and $0 \leq j < 2^l$. A decision unit generates the estimated bit $\hat{u}_i$ which is then broadcast back to every PE. A PE is a configurable element that can perform either function $f$ or $g$. It also includes the $\hat{u}_s$ computation block that updates the $\hat{u}_s$ value with the last decoded bit $\hat{u}_i$ only if the control bit $b_{l,j} = 1$. Another control bit $b_l$ is used to select function $f$ or $g$. Compared to the FFT-like structure, the pipelined tree architecture performs the same amount of computation with the same scheduling (see Table 1) but with a reduced number of PEs and registers. Assuming that a PE (implementing $f$ and $g$) represents twice the complexity of a node processor that implements a single $f$ or $g$ function, the pipelined tree decoder complexity is

$$C_T = (n-1)(2C_{\text{np}} + C_{\text{r}}) + nC_{\text{r}}. \quad (5)$$

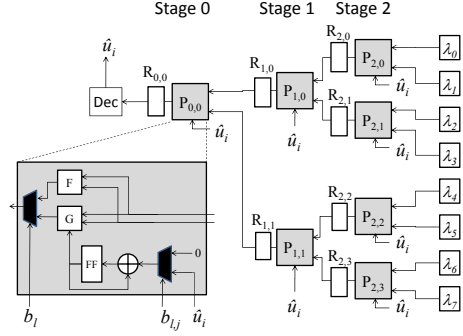

**Fig. 3**. Pipelined tree SC architecture for $n = 8$.

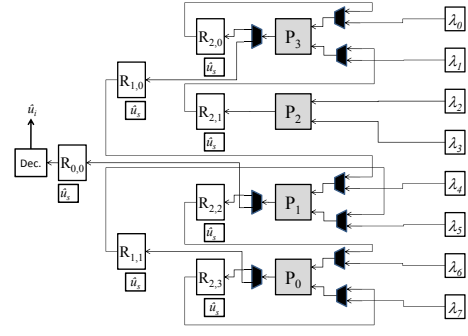

**Fig. 4**. Line SC architecture for $n = 8$.

| CC | 1 | 2 | 3 | 4 | 5 | 6 | 7 | 8 | 9 | 10 | 11 | 12 | 13 | 14 |
|---|---|---|---|---|---|---|---|---|---|---|---|---|---|---|
| $S_2$ | $f$ | | | | | | | $g$ | | | | | | |
| $S_1$ | | $f$ | | $g$ | | | | | $f$ | | | $g$ | | |
| $S_0$ | | | $f$ | $g$ | | $f$ | $g$ | | | $f$ | $g$ | | $f$ | $g$ |
| $\hat{u}_i$ | | | $\hat{u}_0$ | $\hat{u}_1$ | | $\hat{u}_2$ | $\hat{u}_3$ | | | $\hat{u}_4$ | $\hat{u}_5$ | | $\hat{u}_6$ | $\hat{u}_7$ |

**Table 1**. Schedule for the FFT-like and pipeline tree SC architectures ($n = 8$).

Moreover, one can notice that the routing network in the decoder is much simpler in the tree architecture than in the FFT-like structure. Connections between PEs are also local. This lowers the risk of congestion during the wire routing phase of an integrated circuit design and potentially increases the clock frequency and the throughput.

### 3.3. Line SC Architecture

Despite the low complexity of the pipelined tree architecture, it is possible to further reduce the number of PEs. Looking at Table 1, it appears that only one stage is activated at a time. In the worst case (activation of stage $m - 1$), $\frac{n}{2}$ PEs have to be activated at the same time. This means that the same throughput can be achieved with only $\frac{n}{2}$ PEs. The resulting architecture is shown in Figure 4 for $n = 8$. The processing elements $P_j$ are arranged in an line while registers keep a tree structure. Registers and PEs are connected via multiplexing resources that emulate the tree structure. For example since $P_{2,0}$ and $P_{1,0}$ (in Figure 3) are merged to $P_2$ (in Figure 4), $P_2$ should write either to $R_{2,0}$ or $R_{1,0}$ while it should also be able to read from the channel registers or from $R_{2,0}$ and $R_{2,1}$. The $\hat{u}_s$ computation block is moved out of $P_j$ and kept close to the associated register because $\hat{u}_s$ should also be forwarded to the PE. The overall complexity of the line SC architecture is

$$C_T = (n-1)(C_{\text{r}} + C_{\hat{u}_s}) + nC_{\text{np}} + \left(\frac{n}{2} - 1\right) 3C_{\text{mux}} + nC_{\text{r}} \quad (6)$$

where $C_{\text{mux}}$ represents the complexity of a 2-input multiplexer and $C_{\hat{u}_s}$ is the complexity of the $\hat{u}_s$ computation block. Despite the extra multiplexing logic required to route the data through the PE line, the savings in number of PEs makes this SC decoder less complex than the pipelined tree architecture while achieving the same throughput as computed in (4).

It is possible to further reduce the number of PEs with only a small penalty in terms of throughput. Looking at Table 1, during the decoding of one vector, stage $l$ is activated $2^{m-l}$ times. Consequently, in the line architecture of Figure 4, $\frac{n}{2}$ stages are all activated at the same time only twice during the decoding of a vector, regardless of the code size. A decoder with only $\frac{n}{4}$ PEs would require only 2 extra clock cycles to decode a vector. Such a semi-parallel architecture would improve the hardware efficiency at only a small decrease of throughput.

The Line SC architecture can be seen as a tree architecture in which complexity is reduced by merging some of the PEs. An alternative would be to start from the same tree architecture and use the idle stages to overlap the decoding of several codewords at once, enhancing the throughput.

### 3.4. Vector-overlapping SC architecture

Let's assume that we want to use idle cycles in the pipelined tree architecture in order to overlap the decoding of $P$ vectors **y**. At CC #1, $\mathbf{y}_1$ is fed into stage 2 of the pipelined tree decoder. At CC #2, a second vector $\mathbf{y}_2$ is shifted into stage 2 while $\mathbf{y}_1$ uses stage 1. At CC #3, $\mathbf{y}_1$ and $\mathbf{y}_2$ are in stages 0 and 1 respectively. Then, a PE conflict occurs at CC #4 when both $\mathbf{y}_1$ and $\mathbf{y}_2$ need to access stage 0. This problem can be overcome by simply duplicating stage 0 so no resource conflict happens. As shown in Table 2, by duplicating stage 0 (denoted as $S_{0d}$), it is possible to overlap up to 3 vectors at the same time. It would actually be possible to insert another vector by using the remaining idle resources, but the routing of data across the tree would lose its nice regular property, making the multiplexing design more complex. Since several vectors are decoded at the same time, each PE should have access to registers associated with each vector. This means that $P$ register sets are required to decode $P$ vectors in parallel. A vector-overlapping SC decoder is shown in Figure 5 for $n = 8$ and $P = 3$. The degree of parallelism $P$ can actually be enhanced by further duplicating PE stages. It can be shown that in order to reach parallelism $P$, each stage $l$ should be duplicated $\lceil \frac{P+1}{2^{l+1}} \rceil$ times. This vector-overlapping architecture allows us to reach a maximum parallelism value of $P = n - 1$. The complexity and the throughput of a vector-overlapping SC architecture with parallelism $P$ are

$$C_T = \left(n + \frac{P+1}{2}\left[\log\left(\frac{P+1}{2}\right) - 1\right]\right)2C_{\text{np}} + P(2n-1)C_{\text{r}}, \quad (7)$$

$$\text{and} \quad T = \frac{P}{2t_{\text{np}}}. \quad (8)$$

| CC | 1 | 2 | 3 | 4 | 5 | 6 | 7 | 8 | 9 | 10 | 11 | 12 | 13 | 14 | 15 | 16 |
|---|---|---|---|---|---|---|---|---|---|---|---|---|---|---|---|---|
| $S_2$ | $y_1$ | $y_2$ | $y_3$ | | | | | $y_1$ | $y_2$ | $y_3$ | | | | | | |
| $S_1$ | | $y_1$ | $y_2$ | $y_3$ | $y_1$ | $y_2$ | $y_3$ | | $y_1$ | $y_2$ | $y_3$ | $y_1$ | $y_2$ | $y_3$ | | |
| $S_0$ | | | $y_1$ | $y_1$ | $y_2$ | $y_1$ | $y_1$ | $y_2$ | $y_3$ | $y_1$ | $y_1$ | $y_2$ | $y_1$ | $y_1$ | $y_2$ | $y_3$ |
| $S_{0d}$ | | | | $y_2$ | $y_3$ | $y_3$ | $y_2$ | $y_3$ | | | $y_2$ | $y_3$ | $y_3$ | $y_2$ | $y_3$ | |

**Table 2**. Schedule for the vector-overlapping SC architecture ($n = 8$ and $P = 3$).

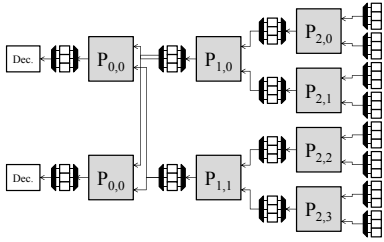

**Fig. 5**. Vector-overlapping SC decoder for $n = 8$ and $P = 3$.

This architecture provides a solution to enhance the parallelism of the decoder without duplicating all the resources of the decoder.

## 4. MINIMUM APPROXIMATION

SC decoding, in its original version, was proposed in the likelihood ratio domain in which the update rules $f$ and $g$ require multiplication and division. The hardware implementation of multipliers and dividers is very expensive and usually avoided in practical decoder designs. We propose to perform SC decoding in the log-domain in order to reduce the complexity of the $f$ and $g$ computation blocks. We assume that the channel information is available as the log-likelihood ratios (LLRs) $L_i$. In the LLR domain $f$ and $g$ become

$$\begin{cases} f(L_a, L_b) = 2\tanh^{-1}\left(\tanh\left(\frac{L_a}{2}\right)\tanh\left(\frac{L_b}{2}\right)\right) \\ g_{\hat{u}_s}(L_a, L_b) = L_a(-1)^{\hat{u}_s} + L_b, \end{cases} \quad (9)$$

where $L_a$ and $L_b$ are LLRs. In terms of hardware implementation, $g$ can be easily mapped to an adder/subtractor controlled by the bit $\hat{u}_s$. However, $f$ involves some transcendental functions that are complex to implement in hardware. One can notice that the $f$ and $g$ functions are identical to the update rules used in BP decoding of LDPC codes. Consequently, similar to what is done in LDPC decoder implementation [6], $f$ can be approximated with the minimum function such that

$$f(L_a, L_b) \approx \text{sign}(L_a)\text{sign}(L_b)\min(|L_a|, |L_b|). \quad (10)$$

| Arch. | $C_{\text{np}}$ | $C_{\text{r}}$ | $T$ |
|---|---|---|---|
| FFT-like | $n \log n$ | $n(1 + \log n)$ | $\frac{1}{2t_{\text{np}}}$ |
| Pipe. Tree | $2n - 2$ | $2n - 1$ | $\frac{1}{2t_{\text{np}}}$ |
| Line | $n$ | $2n - 1$ | $\frac{1}{2t_{\text{np}}}$ |
| Overlap. | $\sim n + \frac{P}{2}(\log \frac{P}{2})$ | $P(2n - 1)$ | $\frac{P}{2t_{\text{np}}}$ |

**Table 3**. Comparison of SC decoder architectures.

In order to estimate the performance degradation incurred by this approximation we simulated the performance of different polar codes on an AWGN channel with BPSK modulation. There was no significant performance loss.

## 5. CONCLUSION

In this paper we showed that the architecture proposed by Arıkan in [1] can be improved by taking advantage of the scheduling in SC decoding. Table 3 is a comparison of the complexity and throughput of the FFT-like SC decoder with the proposed architectures. The pipelined tree architecture and the line architecture allow us to reach the same throughput while reducing the hardware complexity. We also showed that throughput can be enhanced by decoding several vectors in parallel in a vector overlapping architecture.

In this paper, we investigated fully-parallel architectures for SC decoders. For very large code lengths, it would be required to consider semi-parallel architectures in which PEs are shared within the update phase of the same stage as suggested in Section 3.3. The very regular structure of polar codes makes semi-parallel architectures straightforward to implement.

## 6. REFERENCES

[1] E. Arikan, "Channel polarization: A method for constructing capacity-achieving codes for symmetric binary-input memoryless channels," *IEEE Trans. on Inform. Theory*, vol. 55, no. 7, pp. 3051 –3073, Jul. 2009.

[2] H. Mahdavifar and A. Vardy, "Achieving the secrecy capacity of wiretap channels using polar codes," in *IEEE ISIT 2010*, Jun. 2010, pp. 913 –917.

[3] I. Tal and A. Vardy, "How to construct polar codes," in *IEEE ITW 2010*, Aug. 2010.

[4] N. Hussami, R. Urbanke, and S.B. Korada, "Performance of polar codes for channel and source coding," in *IEEE ISIT 2009*, Jun. 2009, pp. 1488 –1492.

[5] E. Arikan, "Polar codes: A pipelined implementation," in *ISBC2010*, Jul. 2010.

[6] M.P.C. Fossorier, M. Mihaljevic, and H. Imai, "Reduced complexity iterative decoding of low-density parity check codes based on belief propagation," *IEEE Trans. on Comm.*, vol. 47, no. 5, pp. 673 –680, May. 1999.